\documentclass[prl,preprintnumbers,amsmath,amssymb,twocolumn]{revtex4}

\usepackage{graphicx}
\usepackage{dcolumn}
\usepackage{bm}
\usepackage{hyperref}
\usepackage{xcolor} 
\usepackage{notoccite} 

\begin{document}


\title{Dendritic signal transmission induced by intracellular charge inhomogeneities} 

\author{Ivan A. Lazarevich}
\author{Victor B. Kazantsev}%

\affiliation{Institute of Applied Physics of Russian Academy of
Science, 46 Uljanov str., 603950 Nizhny Novgorod, Russia }
\affiliation{N.I. Lobachevsky State University of Nizhni Novgorod,
23 Gagarin ave., 603950 Nizhny Novgorod, Russia }

\begin{abstract}
Signal propagation in neuronal dendrites represents the basis for
interneuron communication and information processing in the brain.
Here we take into account charge inhomogeneities arising in the vicinity of ion channels in
cytoplasm and obtained a modified
cable equation. We show that the charge inhomogeneities acting on the
millisecond time scale can lead to the appearance
of propagating waves with wavelengths of hundreds of micrometers.
They correspond to a certain frequency band predicting the appearance of resonant properties in brain neuron signalling.
\end{abstract}

\maketitle

Cable theory is one of the foundations of bioelectrical signal
transmission in nerve tissues \cite{segev,rall,rallexp,wu}. It
describes the propagation of membrane potential in passive
neurites, particularly, in dendrites where the concentration of
active ion channels is not sufficient enough to enable stable
action potential propagation. One of the earliest predictions of the
cable theory was exponential attenuation of postsynaptic
potentials with distance \cite{rall}. However, recent experimental
findings have suggested that distally located synaptic inputs can also
influence the somatic membrane potential \cite{hausser}. 

The cable equation can be derived by constructing an equivalent
electrical circuit with elements describing the electrical properties of neurites \cite{rall}. However, it does not take into
account inhomogeneous distributions of ion concentrations within
dendrites. To overcome this, Qian and Sejnowski \cite{qian}
modified the cable equation by using the Nernst-Planck equation
describing electrodiffusive motion of ions. For spiny dendrites,
several modifications of the cable model were introduced to
account for the influence of the spines on the electrical
characteristics of the cable and anomalously slow diffusion of
ions \cite{henry,koch,baer} as well as active wave propagation
\cite{coombes,timofeeva}. Several authors used the Maxwell's
equations to generalise the cable equation and account for the
charge accumulation nearby endogenous structures of the dendrites
\cite{lindsay,poznanski}. B\'edard and Destexhe \cite{bedard}
considered the non-ideal properties of the membrane as a capacitor
arising from the non-instantaneous motion of ions within the dendrites.

In neuroscience, many recent findings suggested that signal
propagation in the dendritic tree may implement simple information
processing functions at the level of a single neuron
\cite{gutkin2}. Interestingly, that together with the summation of
local inputs attenuated in propagation from periphery to the
active zones in soma resonant and oscillatory properties of the
dendrites were reported \cite{kasevich,gutkin}. These resonances
were typically associated with the presence of active channels in
the dendrites sustaining the propagation.

In this Letter, we show how charge inhomogeneities within the intracellular space can influence passive signal propagation in
dendrites. In particular, we show that the existence of excess
charge areas in the vicinity of ion channels can lead to formation
of travelling waves propagating over larger distances than
solutions predicted by classical cable theory. The model
apparently predicts the appearance of resonant zones in the
frequency band of 25-35 Hz (for the membrane time constant $\tau_m$ = 5 ms) what is in agreement with recent
experimental findings in neuroscience
\cite{gutfreund,leung,sanhueza,pape,chapman}. 

First, let us consider how charge inhomogeneities can be accounted in cable theory. When charge carriers
move in and out of the channel they create regions of {\em excess
charge} in the vicinity of the channel. In this area the potential
is higher than in the surrounding medium \cite{bulai}. The
existence of the overpotential near the channel pore leads to the
increase of the total local potential over the channel. The area
of excess charge can be described as a volume within some closed
surface $E$ covering the channel pore (Fig.1). Therefore, the rate of change of the excess
charge is defined by the difference between the transchannel current and the
lateral relaxation current: 
\begin{equation}
\frac{d Q_e}{d t} = I_e - I_{ch}.
\end{equation}
Because the number of ion channels is high and they are evenly
distributed over the membrane's surface, the overpotential near
the channel end is described by a smooth function $V_e(x,t)$.
The total potential over the channel is $V_{tot} = V_{m} + V_{e}$.
The transchannel current is linearly related to the transchannel potential $I_{ch} = G_{ch} V_{tot}$, where $G_{ch}$ is the conductivity
of the single channel. We suppose
that the excess charge relaxates
with characteristic time $\tau_{\rho}$: $I_{e} = -
{Q_e}/{\tau_{\rho}}$. This time constant
$\tau_{\rho}$ (sometimes called the Maxwell-Wagner time constant
\cite{bedard}) may be estimated by applying the Gauss' law to the
surface $E$. The lateral current density is determined by ${ \bf J
}= \sigma {\bf E} +
\partial {\bf D}/ \partial t$ (where ${\bf E}$ is the electric field, ${\bf D}$ is the displacement field, $\sigma$ is the conductivity of the solution). Then, the value of $\tau_{\rho}$
is given by $\tau_\rho = 2 \varepsilon / \sigma $, where $\varepsilon$ is the permittivity of the solution. 
One can estimate $\tau_{\rho} \preceq $  1 ms, therefore the
excess charge within passive cable relaxates at the millisecond or
sub-millisecond time scale. It is the fastest time scale of the
membrane potential dynamics in neuronal dendrites \cite{softky}.
For the extracellular medium (physiological saline) this time
constant is approximately $10^{-10}$ s, which is much smaller than
the same time-constant on the inner surface of the membrane.
Obviously, the effects of the charge inhomogeneities in the
extracellular medium can be neglected.

In the first approximation, the excess charge is linearly related to the
overpotential, e.g.  $V_e = C_m K_e Q_e$, where
$K_e$ is the so-called channel density factor, $C_m$ is the
membrane capacitance. It follows from numerical solution of the
Nernst-Planck-Poisson problem for cylindrical channel geometry that $K_e$
can be expressed as follows \cite{bulai}:
\begin{equation*}
K_e = \frac{r_{c}^2}{r_{m}^2} \frac{4d+r_c}{r_c} \frac{2r_m -
r_c}{r_m - r_c},
\end{equation*}
\begin{figure}
\includegraphics[scale=0.6]{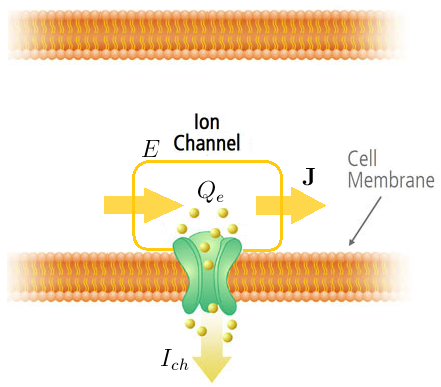}
\caption{Formation of excess charge regions in the vicinity of ion
channels (not to scale). Modified from \url{http://www.neusentis.com/IonChannels.php}}
\end{figure}
where $d$ is the Debye length, $r_c$ is the channel radius, $r_m$
is the size of the membrane patch containing one ion channel.
Under the assumptions that $r_m \gg r_c$ (low channel density) and
$r_c \approx d$ this expression can be simplified to $ K_e = 10
d^2 \pi \xi \sim 10^5 $, where $\xi = 1/\pi r_{m}^2$ is the ion channel density per unit area of the membrane. Combining these assumptions we get the equation for overpotential dynamics:
\begin{equation}
\frac{\partial V_{e} (x,t)}{\partial t} = -\frac{V_e}{\tau_{\rho}}
- \frac{G_{ch}}{C_m K_e}(V_m + V_e).
\end{equation}
Next we apply a set of common assumptions typically used in cable
theory. In particular, we suppose the electric field to be
polarized only in the longitudinal direction ${\bf E}({\bf r}, t)
= E(x,t){\bf x_0}$. Rosenfalk \cite{rosenfalk} and Pickard
\cite{pickard} showed that the magnetic field is negligible
compared to the electric field in neurons due to the slow motion
of charges in the intracellular medium. Hence, the electric field
can be described by a scalar potential $V_m(x,t)$, and $E(x,t) = -
\partial V_m(x,t) / \partial x$. It is assumed that the
extracellular medium can be lumped into a single isopotential
compartment. The intracellular medium is treated as homogeneous
with constant conductivity and the dendritic
segment as being a cylinder with radius $r$. To get the equation
for the membrane voltage, the continuity equation is applied:
\begin{equation*}
\pi r^2 \frac{\partial J(x,t)}{\partial x} + 2\pi r I_{cap}(x,t) +
d_{ch} I_{e}(x,t) = 0
\end{equation*}
Here $I_e(x,t)$ is the current flowing through each excess charge
area with coordinate $x$ at time $t$, $d_{ch}$ is the ion channel
density per unit length, $I_{cap}$ is capacitive current density
per unit length. The total area of the channels within the cable
segment is negligible compared to the segments' area ($r_m \gg
r_c$). Defining $r_i = \sigma / \pi r^2$ and $i_{cap} = 2\pi r
I_{cap}$ we find:
\begin{equation}
\frac{1}{r_i} \frac{\partial^2 V_m(x,t)}{\partial^2 x} = c_m
\frac{\partial V_m}{\partial t} - \frac{d_{ch}C_m
K_e}{\tau_{\rho}} V_e.
\end{equation}
Combining equations (2) and (3), neglecting the terms divided by $K_e
\gg 1$ and noticing that $d_{ch}G_{ch} = g_{ch}$ is the membrane
conductivity per unit length we get
\begin{equation*}
\frac{1}{r_i} \frac{\partial^2 V_m}{\partial x^2} = c_m
\frac{\partial V_m}{\partial t} + g_{ch} V_m - \frac{
\tau_{\rho} }{ r_i }  \frac{\partial^3 V_m}{\partial x^2 \partial
t} + \tau_{\rho} c_m \frac{\partial^2 V_m}{\partial t^2}.
\end{equation*}
Let us introduce membrane time constant, $\tau_m = c_m / g_{ch}$,
and membrane length constant, $\lambda = \sqrt{g_{ch} r_i}$. In
terms of dimensionless variables $X = x/ \lambda$ and $T =
t/\tau_m$ we can write the generalized cable equation in the following
form:
\begin{equation}
\frac{\partial V}{\partial T} + V = \frac{\partial^2
V}{\partial X^2} + \gamma \left( \frac{\partial^3 V}{\partial T
\partial X^2} - \frac{\partial^2 V}{\partial T^2} \right),
\end{equation}
where $\gamma = \tau_{\rho}/\tau_m \ll 1$ is a small parameter.
Note that equation (4) explicitly contains the wave operator $\Box =
\gamma
\partial^2  / {\partial T^2} - {\partial^2 }/{\partial X^2}$.

Performing the Laplace transform of $V(X,T)$
in both space and time we can write:
\begin{equation*}
\hat{V}(k,\omega) = \int_{-\infty}^{+\infty} \int_{0}^{+\infty}
V(X,T) \exp(i\omega T - ikX) dX dT.
\end{equation*}
Substituting it into equation (4) we can express the dispertion relation
in the following form:
\begin{equation}
1 + i\omega = -k^2 - i\gamma k^2\omega + \gamma \omega^2
\end{equation}
An estimate from the Maxwell's equations for typical biophysical
parameters of dendrites gives the value of $\gamma$ of about
$10^{-3}$. Note, however, that larger values of $\gamma$ can be
also considered when finite velocity of charge carriers is taken
into account. B\'edard and Destexhe \cite{bedard}
phenomenologically modified the cable equation to account for
calorific dissipation caused by the charge movement. If we
assume that the excess charge evenly covers the neural membrane
our model will turn into the one obtained in \cite{bedard}. Following this work, we consider the value of $\gamma$
= 0.3. 
Solutions of equation (5) for real wave numbers $k$ and $\gamma$ = 0.3
are presented in Fig. 2. The main difference from the classical
cable model is the emergence of new solutions with $ \text{Re}
\omega \neq$ 0. This means that for a certain range of frequencies
there exist travelling waves which decay with characteristic time
given by $1 / \text{Im} \omega$. Note that for the value of
$\gamma$ = 0.3 the real part of $ \omega $ is non-zero as $k
\rightarrow 0$, which means that the phase velocity of the wave
tends to infinity. Let $\omega = \omega' + i\omega''$ ($\omega',\omega'' \in \mathbb{R}$). The interval of wavenumbers where $\omega' \neq 0$ is given by
\begin{equation*}
k_* = \frac{\sqrt{1-2\sqrt{\gamma}}}{\sqrt{\gamma}} < k < k^* = \frac{\sqrt{1+2\sqrt{\gamma}}}{\sqrt{\gamma}}, \hspace{0.1in} \gamma < 0.25  
\end{equation*}
\begin{equation*}
k_* = 0 < k < k^* = \frac{\sqrt{1+2\sqrt{\gamma}}}{\sqrt{\gamma}}, \hspace{0.2in} \gamma > 0.25
\end{equation*}
These conditions define the \emph{oscillatory zone}, which size is equal to $\Delta k = k^* - k_* = 2 + \mathcal{O}(\sqrt{\gamma})$. If $k$ belongs to the oscillatory zone, the frequency $\omega$ is given by
\begin{figure}
\includegraphics[scale=0.53]{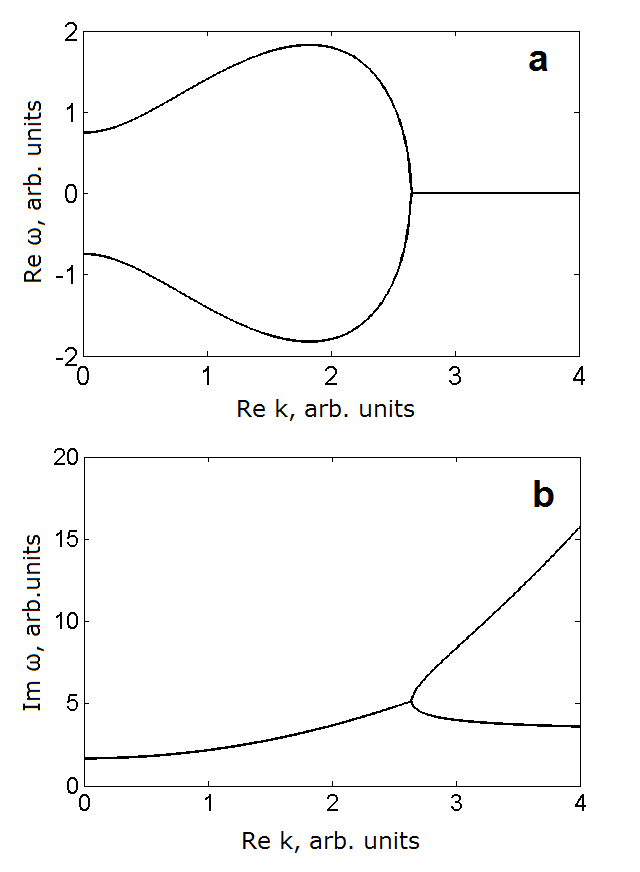}
\caption{Dispertion curves of the modified cable equation. Shown are the (a) real and (b) imaginary parts of
$\omega$ (in units of $1/\tau_m$) for real $k$ (in units of
$1/\lambda$) and $\gamma = 0.3$. There is apparently a frequency
band for which travelling waves exist.
}
\end{figure}
\begin{equation*}
\omega'' = \frac{1}{2\gamma} + \frac{k^2}{2}, \hspace{0.35in} \gamma \omega'^2 = -\frac{1}{4\gamma}(1-\gamma k^2)^2 + 1
\end{equation*}
which implies that in the oscillatory zone the frequency $\omega'$ varies from $0$ to $\omega'_{max} = 1/\sqrt{\gamma}$. Outside the oscillatory zone we have the following two solutions:
\begin{equation*}
\omega''_{\pm} = \frac{1}{2\gamma}\left((1+\gamma k^2) \pm \sqrt{(1-\gamma k^2)^2-4\gamma}\right)
\end{equation*}
For such travelling waves the effective propagation distance, e.g.
the distance at which the signal amplitude decreases by a factor
$1/e$, is $L_{\text{prop}} = (1/k) \text{Re} \omega /\text{Im}
\omega$. The dependence of $L_{\text{prop}}$ on the signal
frequency is presented in Fig. 3. It illustrates that there exist a range of
frequencies defining the \emph{resonant zone} (25-35 Hz for $\tau_m$ = 5 ms), where the travelling waves propagate over larger distances
than the ``diffusing'' solutions of the classical cable model (dotted curve in Fig.3). Moreover, there is a resonant frequency given by
$f_{cr} = \sqrt{4\gamma - 1} / 4\pi\gamma \tau_m \approx 25$ Hz
for which the propagation distance theoretically tends to
infinity as 
\begin{equation*}
L_{\mathrm{prop}} \sim \frac{\omega' \sqrt{\gamma} }{\sqrt{\omega_{cr}(\omega'-\omega_{cr})}}
\end{equation*}
Our formalism can be also applied to a boundary-condition problem.
Consider a finite dendrite of electrotonic length $L$. Suppose that on the left
end of the dendrite ($X=0$) there is a voltage oscillation with
the frequency $\omega'$. It is formulated by the following
boundary conditions:
\begin{equation*}
\left. V \right|_{X=0} = V_{\omega} \exp(i \omega' T),
\hspace{0.2in} \left. \frac{ \partial V}{\partial X}
\right|_{X=L} = I^{L}(T)
\end{equation*}
where the frequency $\omega'$ is taken so that the travelling wave
solutions may exist, and $I^{L}(T)$ is an arbitary function. The travelling wave solutions have the following
form
\begin{figure}
\includegraphics[scale=0.4]{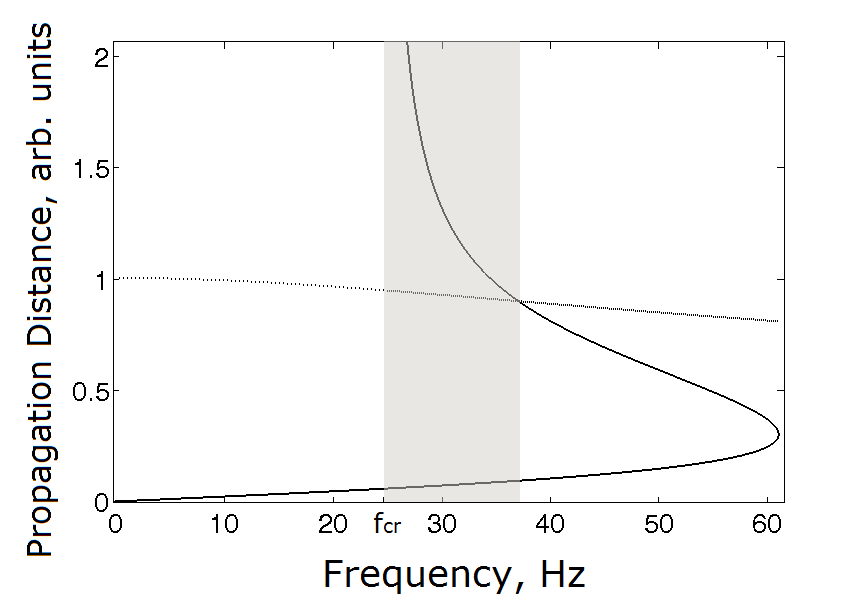}
\caption{Comparison of propagation distances for the standard and modified cable models. Shown are the propagation distances $L_{\text{prop}}$ of travelling wave
solutions (solid line) and standard cable model solutions (dotted
line) in length constants $\lambda$ vs. signal frequency (in Hz). The grey area shows the resonant zone, in which the travelling waves propagate over larger distances than cable model solutions. Propagation distance of cable model solutions is given by $L_{cab} = $ (Im$\sqrt{-1-i\omega'})^{-1}$. The membrane time constant is set to $\tau_m =$ 5 ms.}
\end{figure}
\begin{equation*}
V_{\sim}(X,T) = A \exp(-\omega''T) \exp(i \omega' T-ikX)
\end{equation*}
Any solution of the standard cable equation $V_0(X,T)$ will also
be an approximate solution to the generalized cable equation:
\begin{equation*}
\gamma \left| \left( \frac{\partial^3 V_0}{\partial
T \partial X^2} - \frac{\partial^2 V_0}{\partial T^2} \right)
\right| = \gamma \left| \frac{\partial V_0}{\partial T} \right| \ll \left|
\frac{\partial V_0}{\partial T} \right|
\end{equation*}
Consider the superposition of the solutions $V(X,T) = V_{\sim} +
V_{0}$ which will also be a solution to the Eq. (4) as long as Eq.
(4) is linear. Rewriting the boundary conditions in terms of the
unknown function $V_0(X,T)$, we get
\begin{eqnarray*}
\left. V_0 \right|_{X=0} = -\left. V_{\sim} \right|_{X=0} + V_{\omega} \exp(i \omega' T), \\
\left. \partial_{X} V_0 \right|_{X=L} =  - \left.
\partial_{X} V_{\sim} \right|_{X=L} + I^{L}(T)
\end{eqnarray*}
We also set the initial conditions for $V_0(X,T)$ to be $V_0(X,0)
= - V_{\sim}(X,0)$. Thus, the function $V_0(X,T)$ is now uniquely
defined by one initial and two boundary conditions. Rewriting the
cable equation for $V_0(X,T)$ in terms of $V(X,T)$ we find
\begin{equation*}
V + \frac{\partial V}{\partial T} = \frac{\partial^2 V}{\partial X^2} + \alpha V_{\sim} 
\end{equation*}
where $\alpha = k^2 + i(\omega' + i\omega'') + 1$. Thus, for a set of solutions with particular
initial conditions, Eq. (4) is reduced to the standard cable
equation with a definite source term. The quantity $I_{\text{eff}}
= - \alpha g_{ch} V_{\sim}$ can be interpreted as an effective
membrane current density arising due to resonant properties of the
dendritic tissue. Let us consider the zero initial
condition, $V(X,0) = 0$. Matching the initial condition for
$V_T$ with the boundary conditions we find $ A = i\omega'
V_\omega / \alpha $ and $- i k \alpha A \exp(-ikL) = I_{T}^{L}(T =
0) $, which for small $k \ll 1$ implies $I_{T}^{L}(T = 0) \ll 1$. In case of a sealed end ($I^L(T) = 0$) one can obtain the following constraint on $k$ to satisfy Re$I_T^L(0) = 0$:
\begin{equation*}
kL = \text{Arg}\alpha(k) + \pi n, \hspace{0.1in} n \in \mathbb{Z}
\end{equation*}
which is a transcendental equation determining the possible wavenumbers $k(L)$. 

To summarize our results, we have derived a modified cable equaiton taking into account the finite velocity of charge carriers within the intracellular dendritic space. The model accounts for the excess
charge regions in the vicinity of
intracellular structures such as ion channels of the neuron membrane. A mathematical
derivation of the equation governing voltage in a one-dimensional
cable with an intrinsic inhomogeneous charge distribution is
presented. The modified equation represents a linear cable equation with
additional terms arising due to the overpotential induced by the inhomogeneous distribution of charge carriers in the dendrites.
Our model predicts the existence of a resonance frequency band for which electrical oscillations observed in dendrites
may propagate as travelling waves with relatively large
wavelenghts (several length constants, i.e. hundreds of
micrometers). The critical resonant frequency depends only on the
membrane time constant and on the characteristic time of charge
relaxation. In particular, it does not depend on the radius of the dendritic segment. Our results also suggest that purely passive
dendrites may exhibit resonant properties typically associated
with the presence of active ion channels.


\begin{thebibliography}{9}

\bibitem{segev}
Segev, I. \& London, M.
Untangling dendrites using quantitative models.
\emph{Science} {\bf 290} (5492), 744-750, (2000).

\bibitem{rall}
  Rall, W.,
  \emph{The Theoretical Foundation of Dendritic Function}.
  (MIT Press, Cambridge, MA,
  1995).

\bibitem{rallexp}
Rall, W. 
Branching dendritic trees and motoneuron membrane resistivity. 
\emph{Exp. Neurol.} {\bf 1}, 491-527, (1959).

\bibitem{wu}
  Jonston, D \& Wu, S. M.
  \emph{Foundations of Cellular Neurophysiology}.
  (MIT Press, Cambridge, MA,
  1995).

\bibitem{hausser}
  Hausser, M.
  Synaptic function: dendritic democracy.
  \emph{Curr. Biol.} {\bf 11}(1) (2001).

\bibitem{qian}
Qian, N. \& Sejnowki, T. J.
An electro-diffusion model for computing membrane potentials and ionic concentrations in branching dendrites, spines and axons.
\emph{Biol. Cybern.}, {\bf 62}, 1-15, (1989).

\bibitem{henry}
Henry, B. I., Langlands, T. A. M., Wearne, S. L.
Fractional cable models for spiny neuronal dendrites. 
\emph{Phys. Rev. Lett.} {\bf 100}(12), 128103, (2008).

\bibitem{koch}
Koch, C. 
\emph{Biophysics of Computation, Information Processing in Single Neurons, Computational Neuroscience}
(Oxford University, New York, 1999)

\bibitem{baer}
Baer S. M. \& Rinzel J. 
Propagation of dendritic spikes mediated by excitable spines: a continuum theory. 
\emph{J. Neurophysiol.} {\bf 65}(4), 874-90, (1991)

\bibitem{coombes}
Coombes, S. \& Bressloff, P. C. 
Saltatory waves in the spike-diffuse-spike model of active dendritic spines,
\emph{Phys. Rev. Lett.}, {\bf 91}, 028102, (2003)

\bibitem{timofeeva}
Timofeeva, Y., Cox, S. J., Coombes, S., Josic, K.
Democratization in a passive dendritic tree: an analytical investigation, 
\emph{J. Comp. Neurosci.}, {\bf 25}, 228-244, (2008).

\bibitem{lindsay}
Lindsay, K. A., Rosenberg, J. R., Tucker, G.
From Maxwell's equations to the cable equation and beyond.
\emph{Prog. Biophys. Mol. Biol.}, {\bf 85}(1), 71-116, (2004).

\bibitem{poznanski}
  Poznanski, R. R. 
  Thermal noise due to surface-charge effects within the Debye layer of endogenous structures in dendrites.
  \emph{Phys. Rev. E} {\bf 81}(2), 021902, (2010).

\bibitem{bedard}
  B\'edard, C. \& Destexhe, A.
  A modified cable formalism for modeling neuronal membranes at high frequencies.
  \emph{Biophys. J.} {\bf 94}(3) (2008).

\bibitem{gutkin2}
Caze, R.D., Humphries, M., Gutkin, B. 
Passive dendrites enable single neurons to compute linearly non-separable functions. 
\emph{PLoS Comput. Biol.}, {\bf 9}(2), e1002867 (2013).

\bibitem{kasevich}
  Kasevich, R. S. \& LaBerge, D. 
  Theory of electric resonance in the neocortical apical dendrite.
  \emph{PLoS ONE} {\bf 6}(8), e23412, (2011).

\bibitem{gutkin}
  Remme, M. W. H., Lengyel, M., Gutkin, B. S. 
  The role of ongoing dendritic oscillations in single-neuron dynamics. 
  \emph{PLoS Comput. Biol.} {\bf 5}(9), e1000493, (2009).
    
\bibitem{gutfreund}
  Gutfreund, Y., Yarom, Y., Segev, I. 
  Subthreshold oscillations and resonant frequency in guineapig cortical neurons: physiology and modelling. 
  \emph{J. Physiol.} { \bf 483} (Pt 3), 621–640, (1995)
  
\bibitem{sanhueza}
  Sanhueza, M. \& Bacigalupo, J. 
  Intrinsic subthreshold oscillations of the membrane potential in pyramidal neurons of the olfactory amygdala.
\emph{Eur. J. Neurosci.} {\bf 22}, 1618–26, (2005).
  
\bibitem{pape}
  Pape, H. C., Pare´, D., Driesang, R. B. 
  Two types of intrinsic oscillations in neurons of the lateral and basolateral nuclei of the amygdala. 
  \emph{J. Neurophysiol.} {\bf 79}, 205–16, (1998).

\bibitem{leung}
Leung, L. W. \& Yim, C. Y. 
Intrinsic membrane potential oscillations in hippocampal neurons in vitro. 
\emph{Brain Res.}, { \bf 553}, 261–74, (1991).

\bibitem{chapman}
Chapman, C. A. \& Lacaille, J. C. 
Intrinsic theta-frequency membrane potential oscillations in hippocampal CA1 interneurons of stratum lacunosum-moleculare.
\emph{J. Neurophysiol.}, {\bf 81}, 1296–307, (1999).

\bibitem{bulai}
  Bulai, P. M. et. al
Extracellular electrical signals in a neuron-surface junction: model of heterogeneous membrane conductivity.
  \emph{Eur. Biophys. J.} {\bf 41}(3), 319-327 (2012).  
  
\bibitem{softky}
Softky, W.
Sub-millisecond coincidence detection in active dendritic trees.
\emph{Neurosci.} { \bf 58}, 15-41, (1994).

\bibitem{rosenfalk}
  Rosenfalck, P. 
  Intra- and extracellular potential fields of active nerve and muscle fibers. 
  \emph{Acta Physiol. Scand. Suppl.} {\bf 47}, 239–246, (1969).
    
\bibitem{pickard}
  Pickard, W. F. 
  The electromagnetic theory of electrotonus along an unmyelinated axon. 
  \emph{Math. Biosci.} {\bf 5}, 471-494, (1969).

\end{thebibliography}
\end{document}